\documentstyle[12pt,epsfig]{article}

\catcode`@=12 \topmargin -0.5in \evensidemargin 0.0in
\newcommand{\beq}{\begin{equation}}
\newcommand{\eeq}{\end{equation}}

\newcommand{\bea}{\begin{eqnarray}}
\newcommand{\eea}{\end{eqnarray}}

\catcode`@=12 \topmargin -0.5in \evensidemargin 0.0in
\begin{document}
\thispagestyle{empty}
\hbox{}
\vspace{1.in}
\begin{center}
{\Large \bf Muon anomalous magnetic moment in models with singlet fermions}\\
 \vspace{1in} {Subhash Rajpoot\\
Department of Physics and Astronomy\\ California State University\\
Long Beach, CA 90840, USA}  \vspace{1.in}
\end{center}
\begin{abstract}
It is shown that a minimal extension of the  standard model with
 leptons that transform as singlets under the $SU(2)$
 symmetry of weak interactions can explain the recently reported
derivation in the measured value of the
anomalous magnetic moment of the muon from that expected in the standard
model.

\end{abstract}

\newpage
\baselineskip 24pt

Recently  the Muon $(g-2)$ Collaboration \cite{bro} has reported a
measurement of the anomalous
 magnetic moment of the positively charged muon,
\begin{equation}
a_{\mu}^{exp}=11 659 202(14)(6) \pm 165 \cdot 10^{-11}. \end{equation}
The value expected in the standard model is
\begin{equation}
a_{\mu}^{SM}= 11 659 159.6(6.7)\cdot 10^{-11}
\end{equation}

There seems to emerged a discrepancy of
2.6 $\sigma$ deviation between theory and the world averaged experimental
value
\begin{equation}
a_{\mu}^{exp}-a_{\mu}^{SM}=426 \pm 165 \cdot 10^{-11}.
\end{equation}

Moreover, at 90$\%$ confidence level, 
the measurement indicates a relatively large effect
due to "new physics" beyond the standard model. 
Since  the value tends to lie in
the
"positive" rather than in the "negative" direction, it would appear that the
discrepancy
can only
accommodate
  a very selective type
of new physics.
To the contrary, the discrepancy allows for several
 possible scenarios \cite{mar} to be entertained;  supersymmetry
\cite{sus}, scalar leptoquarks
\cite{slq}, substructure\cite{ss},  muon substructure \cite{kan}, new gauge
bosons\cite{ngb}, exotic
fermions \cite{ef} and models for neutrino masses\cite{ma}. Here,
  a minimal extension of the standard model is proposed which contains
fermions transform as
singlets under the $SU(2)$ of electroweak interactions.

One way to motivate the extension of the standard model with singlet
fermions
is the  generalized see-saw mechanism \cite{raj},
widely discussed
in the context of neutrino masses. The generalized see-saw mechanism
explains in similar
terms the smallness of the masses of the fermions belonging to
at least the first two generations.
 Here, to keep matters simple, we propose to extend the
second family with singlet charged leptons $M_L$ and $M_R$  that are
vectorlike , i.e.,
they transform as
\begin{equation}
M_L \sim (1,1,-2),\,\,\,\,\,\,\,\,\, M_R \sim (1,1,-2),
\end{equation}
under the standard model gauge group  $SU(3) \times SU(2) \times U(1)$.
The model still contains the conventional quarks and leptons
 and a single doublet of Higgs and is  free from triangle
anomalies.
The singlet fermions may have a
bare mass term since no symmetry forbids it.

The coupling of the fermions to the muons are given by
\begin{equation}
L = Y_{\mu M}(\bar\nu_{\mu} \; {\bar\mu} )_L 
i\sigma_2\Phi^*M_R
+ m_{\mu M}\bar{M_L}\mu_R + {\mbox{(H.C)}},
\end{equation}
where $Y_{\mu M}$ and $m_{\mu M}$ refer to Yukawa couplings. Without loss of
generality we take $Y_{\mu M}$ equal to $m_{\mu M}$.
After spontaneous
symmetry breaking, there is left over the usual Higgs boson $h^0$ with
coupling $Y_{\mu M}$
 to the $\mu$ and $M$.

The contribution of the singlet heavy charged lepton $M$
to $a_{\mu}$ is due to its couplings to the neutral Higgs
 boson and the muon as shown in Fig.1.
This contribution is  given by \cite{lev}
\begin{equation}
 a_{\mu}={Y_{\mu M}^2 m^2_{\mu}\over 8\pi^2} \int_0^1
{x^2(1-x) +  x^2{m_{M} / m_\mu} \over
x(x-1)m^2_{\mu}+x{m_{M}^2}+(1-x){m^2_h}}\ dx.
\end{equation}

Here $a_{\mu}$ is defined as the coefficient of the term
\begin{equation}
ie\frac{\sigma_{\mu\nu}q^{\nu}}{4m_{\mu}}
\end{equation}
in the effective Lagrangian. There are additional 
contributions with standard model gauge boson exchange
but they are known to be small.
In evaluating $a_\mu$, we take the Higgs boson to be light with mass
 $m_{H}=100$~GeV, which is
compatible with the present bounds from phenomenology. In Fig.2, $a_{\mu}$
is plotted as a function of $m_M$ varying
 between 0.1 TeV and 5 TeV for three different values of $Y_{\mu M}$.
 At 90$\%$ confidence level
$a_{\mu}$ lies between
\begin{equation}
21.5 \times 10^{-10}   \le  a_{\mu}   \le 63.7 \times  10^{-10}.
\end{equation}
As can bee seen from  Fig.2 these bounds select the value of $Y_{\mu M}=0.1$ 
and the following bounds on the mass of the singly charged lepton
\begin{equation}
{3 \,\mbox {TeV}}  \ge  m_M   \ge {1 \,\,\mbox {TeV}}.
\end{equation}

The present limits on the mass of singly charged
 heavy lepton are $m_{M}\sim 95$ GeV \cite{pdg}.
 Thus, with  $Y_{\mu M}\sim 0.1$
and $m_{M}$ of order 1 TeV, the
anomaly can easily be explained. It is to be noted that the addition of
the singlet lepton leads to flavor changing processes such
 as $\mu\rightarrow e\gamma$. However, the predicted rate is model dependent
and even if we make the conservative estimate of taking $Y_{\mu M}=Y_{e M}$,
the rate is well below the
 bounds deduced from present experiments.

To conclude, if the  reported anomaly \cite{bro} in the theoretical and
experimental
values of the muon anomalous
magnetic moment of the muon withstands the test of time \cite{ynd},
 then a simple viable explanation 
can be the standard model with an extended second generation of
 conventional
fermions with weak $SU(2)$ singlets charged leptons.
 The singlet
leptons required in the scheme lie within the reach of the near future
upcoming accelerators.

\begin{figure}
\begin{center}
\psfig{file=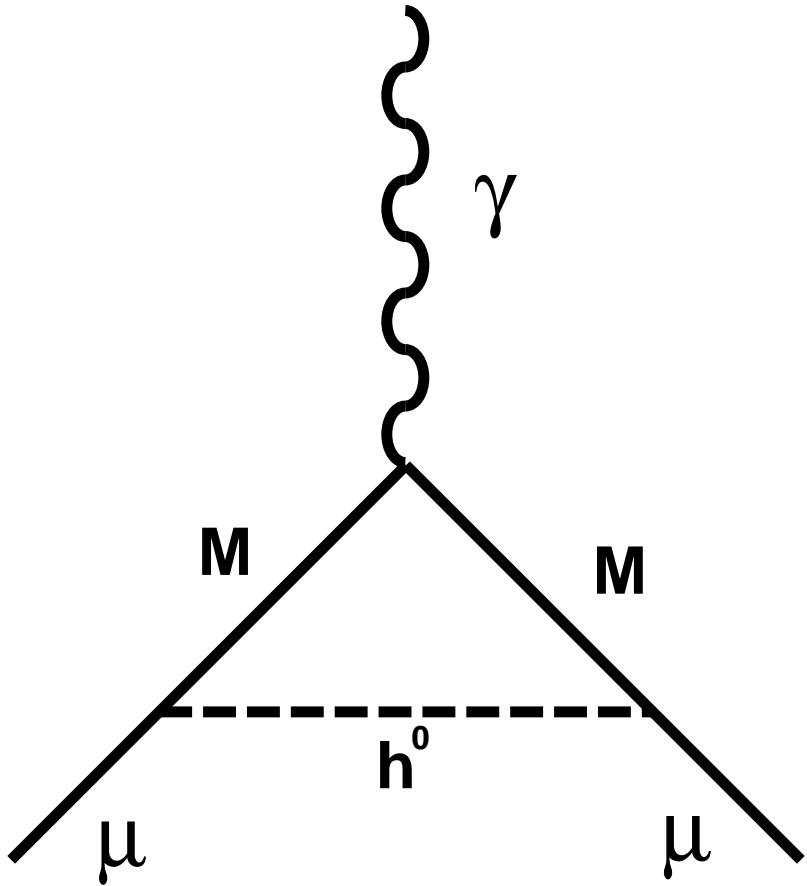,width=10.5cm}
\end{center}
\caption{Singlet fermion contribution to muon anomalous magnetic moment.}
\label{fig1}
\end{figure}

\begin{figure}
\begin{center}
\psfig{file=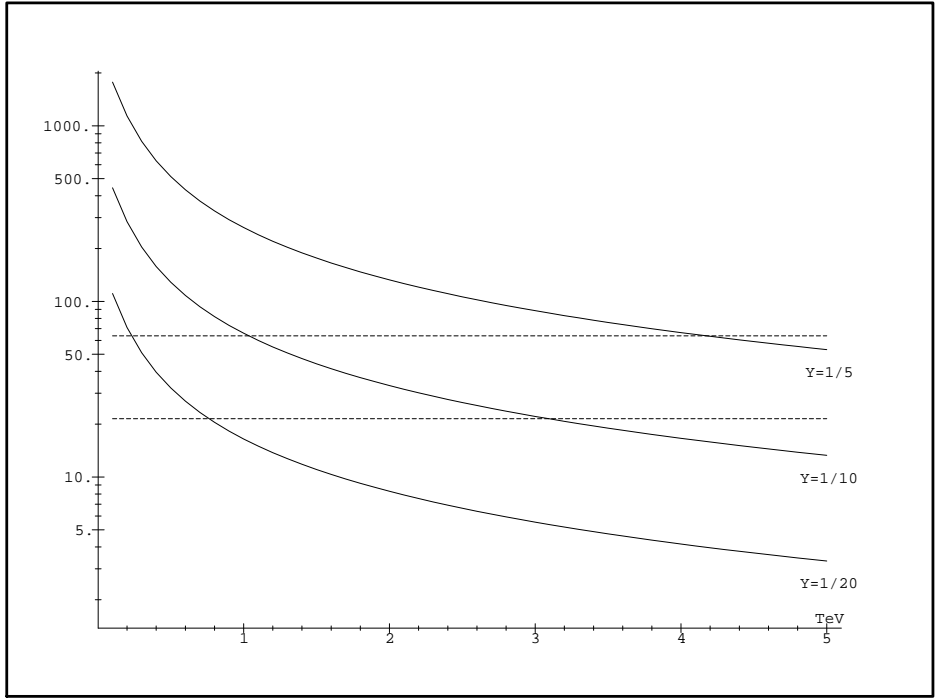,width=10.5cm,angle=-90}
\end{center}
\caption{Muon magnetic moment $a_{\mu M}$ (multiplied by $10^{10}$) 
as a function of $Y_{\mu M}$ and
singlet lepton mass $m_M$. }
\label{fig2}
\end{figure}

\end{document}